# An Antarctic ice core recording both supernovae and solar cycles

Yuko Motizuki[1], Kazuya Takahashi[1], Kazuo Makishima[2,3], Aya Bamba[2,†], Yoichi Nakai[1], Yasushige Yano[1], Makoto Igarashi[2], Hideaki Motoyama[4], Kokichi Kamiyama[4], Keisuke Suzuki[5], Takashi Imamura[6]

*[1]RIKEN Nishina Center, Hirosawa 2-1, Wako 351-0198, Japan*

*[2]Cosmic Radiation Laboratory, RIKEN, Hirosawa 2-1, Wako 351-0198, Japan*

*[3]Department of Physics, University of Tokyo, Hongo 7-3-1, Bunkyo-ku, Tokyo 113-0033, Japan*

*[4]National Institute of Polar Research, Kaga 1-9-10, Itabashi-ku, Tokyo 173-8515, Japan*

*[5]Department of Environmental Sciences, Shinshu University, Asahi 3-1-1, Matsumoto 390-8621, Japan*

*[6]National Institute for Environmental Studies, Onogawa 16-2, Tsukuba 305-8506, Japan*

*[†]Present address: ISAS/JAXA, Yoshino-dai 3-1-1, Sagamihara 229-8510, Japan*

Ice cores are known to be rich in information regarding past climates, and the possibility that they record astronomical phenomena has also been discussed. Rood *et al.*[1] were the first to suggest, in 1979, that nitrate ion ($NO_3^-$) concentration spikes observed in the depth profile of a South Pole ice core might correlate with the known historical supernovae (SNe), Tycho (AD 1572), Kepler (AD 1604), and SN 1181 (AD 1181). Their findings, however, were not supported by subsequent examinations by different groups using different ice cores[2–4], and the results have remained controversial and confusing[5,6]. Here we present a precision analysis of an ice core drilled in 2001 at Dome Fuji station in Antarctica. It revealed highly significant three $NO_3^-$ spikes dating from the 10th to the 11th century. Two of them are coincident with SN 1006 (AD 1006) and the Crab Nebula SN (AD 1054), within the uncertainty of our absolute dating based on known volcanic signals. Moreover, by applying time-series analyses to the measured $NO_3^-$ concentration variations, we discovered very clear evidence of an 11-year periodicity that can be explained by solar modulation. This is one of the first times that a distinct 11-year solar cycle has been observed for a period before the landmark studies of sunspots by Galileo Galilei with his telescope. These findings have significant consequences for the dating of ice cores and galactic SN and solar activity histories.

When a SN explodes close to the earth, the nuclear γ-rays produced (~0.1–1 MeV) can lead to an enhancement of nitrogen oxide production in the middle atmosphere or the stratosphere. SN events could therefore be traced in ice cores as $NO_3^-$ concentration spikes. Possible SN-caused $NO_3^-$ concentration variations in Antarctica, however, can be disturbed by nitrogen oxides transported thorough the troposphere from lower latitudes. Since such nitrogen oxides tend to precipitate in coastal regions, drilling sites in inland parts of Antarctica are preferable for our purpose. Dome Fuji station (Fig. 1) is an inland site, and hence ice cores obtained there are likely to contain much stratospheric information. The 10 m depth snow temperature at Dome Fuji is –57.3 ºC, and the mean accumulation rate (MAR) of ice from 1995 to 2006 was 27.3±1.5 mm water-equivalent $yr^{-1}$ [7].

Bombardment of the atmosphere by high-energy (~10–500 MeV) protons originating from solar flares, so-called solar proton events (SPEs), may also leave $NO_3^-$ concentration spikes in ice cores[8]. Consequently, we must be careful to consider possible SPEs when we search ice cores for SN signals, and must carefully choose the era such that the core samples are less likely to be affected by SPEs. The period between AD ~1040 and ~1060 is known to be quiet in auroral activity and, hence, solar activity[9]. Accordingly, probably no large solar flares accompanied by SPEs occurred in these two decades. Coincidentally, the SN explosion in AD 1054 of the Crab Nebula occurred during this time period, making it a good target, as recommended by Stothers[10]. We therefore analysed an ice core that included ice deposited during the 11th century, searching for a pair of $NO_3^-$ peaks that might be associated with the two historical SN events in that century: SN 1006 and the Crab Nebula. A pair of $NO_3^-$ spikes, separated by ~50 years, would strongly support a possible correlation between SNe and $NO_3^-$ concentration spikes.

We studied a potion of a 122-m-long ice core drilled in 2001 (hereafter referred to as the DF2001 core). The core was cut at Dome Fuji and transported to the National Institute of Polar Research (NIPR, Japan) in 50-cm-long segments. These segments were further cut into shorter pieces; those from core depths below 50 m were each ~3 cm long and those from shallower depths were ~4 cm long. We thereby obtained a total of 323 samples covering approximately 200 years, with an average time resolution of 0.6 year per sample. Using high-performance ion chromatography, we analysed these samples for anion and cation concentrations. The typical measurement precision for $NO_3^-$ concentration was within ±0.01 μeq L$^{-1}$ as detailed in Supplementary Information. Figure 2 shows the $NO_3^-$ concentration profile from 45.5 to 56.8 m depth. The results revealed two sharp $NO_3^-$ concentration spikes at depths of around 47.40 and 48.07 m, and another broader spike at around 50.85 m. We performed replicate measurements on the residuals of the samples yielding the spikes, and successfully reconfirmed (within error) these results (Fig. 2).

Our overall $NO_3^-$ measurement data gave a mean of 0.247 μeq L$^{-1}$ and a standard deviation (1σ) of 0.057 μeq L$^{-1}$. The spikes at 47.40, 48.07, and 50.85 m depth were accordingly found to deviate from the mean by 3.6σ (two data points), 6.7σ (one data point), and 5.1σ (two data points), respectively. Although the histogram of all the data was rather asymmetric, if the data distribution is assumed to be Gaussian with these mean and 1σ values, the statistical probability that a data point would fall outside the 3.6σ threshold is 0.032%. We should thus expect to find only $323 \times 3.2 \times 10^{-4} = 0.1$ such cases among the 323 analysed samples. It is therefore clear that the three spikes are highly significant and cannot be explained by mere statistical fluctuations.

Such large fluctuations might originate from contamination artefacts. In particular, such artefacts are more likely at the edges of a core segment[11]. In our study, the spikes at around 47.40 and 50.85 m depth were from several samples from the middle of core segments, but that at 48.07 m was measured on a sample at a segment end. We observed, however, that this core segment was clean, and its connection to the adjacent segment was without defect. Furthermore, the average $NO_3^-$ concentration in our core-end samples, $0.243 \pm 0.060$ μeq $L^{-1}$, exhibits no excess above the mean, $0.247 \pm 0.057$ μeq $L^{-1}$, averaged over the entire 323 samples (both errors are 1σ values). Hence, no contamination was found in our core-end samples. We further confirmed that the three spikes were not accompanied by any sudden changes in water isotope ($\delta^{18}O$, $\delta D$, d-excess) concentrations, thus excluding the possibility that any sudden violent climate changes occurred at those times. We therefore conclude that the three $NO_3^-$ concentration spikes are real, and reflect natural phenomena other than climatic variation.

The depth–age relation in ice cores is usually determined by using past volcanic eruption signals as absolute time markers, and the portion of the core between two adjacent time markers is dated by assuming a constant MAR between them[12]. We detected clear non-sea-salt sulphate ($nssSO_4^{2-}$) spikes at 38.80 and 85.61 m depth, which can be associated with known volcanic eruptions in AD 1260 (possibly of El Chichon, Mexico[13]; assuming a 1-year lag between the eruption date and the signal in the core[12]) and AD 180 (Taupo, New Zealand[14]), respectively. We also found two other major peaks in the nssSO$_4^{2-}$ profile, at depths of 42.48 and 70.29 m, and identified them with volcanic eruptions in AD 1168[15,16] and AD 639[16,17], respectively. These events have been detected in bipolar ice core records, and the dates are considered accurate to within ±5 years[15] for the AD 1168 event and ±2 years[17] for the AD 639 event. The depth of the AD 1168 event in our DF2001 core, relative to the well-established AD 1260 peak, is consistent with that predicted by the MAR between AD 1260 and 1454 (26.4 mm water-equivalent yr$^{-1}$). The AD 639 event was also identified in another core, drilled in 1993 at Dome Fuji station[18], and the depth agrees well with the present identification.

These identifications of volcanic eruptions result in MARs of 25.1, 35.5, and 24.3 mm water-equivalent yr$^{-1}$ for AD 180–639, 639–1168, and 1168–1260, respectively (Figure 3, model A). The clear enhancement of MAR between 42.48 and 70.29 m depth (AD 1168 and 639, respectively; see Fig. 3) strongly suggests the existence of the so-called Medieval Climatic Anomaly (MCA)[19]. In fact, large increases in annual layer thicknesses were observed[20] at around 50 m depth in the core obtained at Dome Fuji in 1993; sudden decreases in the $\delta^{18}$O profile were also found at the depths where the annual layer thickness increased dramatically.

At the Dome Fuji site, the MCA probably occurred between the identified volcanic events of AD 639 and 1168 (*i.e.*, within this 529-year interval), because the dates of the MCA are usually recognized as between AD 800 and 1300[19], and because its duration has been suggested to be ~400 years in a core from the South Pole[21] and also in one from near the Antarctic Peninsula[22]. In a Greenland core, the MCA, associated with a large MAR, has been reported to have lasted from AD ~700 to ~1100[23]. On the basis of these findings, we infer that the MCA was recorded from 69.10 to 45.38 m depth, where strong decreases occur in the $\delta^{18}$O depth profile[24] in the DF2001 core. We then assumed that the MAR calculated for the AD 180–639 interval lasted until the MCA starting depth (69.10 m), and that the MAR calculated for AD 1168–1260 began at the MCA ending depth of 45.38 m (Fig. 3, model B). Accordingly, we calculated the MCA starting and the ending depths in model B to be AD 673 and 1094, respectively (Fig. 3). The resulting MCA duration, ~420 years, is consistent with the reported values for Antarctica[21,22], described above. Note that our data region of interest was totally within the MCA period of either model A or B (Fig. 3). The model B MAR during the MCA was 38.3 mm water-equivalent yr$^{-1}$ (Fig. 3). If the MCA duration were much shorter than the model B duration, the deduced MAR during the MCA would be much larger than that indicated by the water isotope variations obtained so far from the Dome Fuji ice cores.

At the top of Figure 2, we have plotted our model chronologies A and B, determined by using model MARs A and B, respectively, shown in Figure 3. As a result, two of the three spikes fall very close to the explosion dates of SN 1006 and the Crab Nebula (AD 1054). By assuming a 1-year lag between each explosion date and the signal in the ice core, we can estimate the associated date uncertainty from the model chronologies. Accordingly, the estimated uncertainty ranges of the absolute ages corresponding to AD 1007 (SN 1006) and AD 1055 (the Crab Nebula) are 15 and 19 years, respectively (horizontal bars in Fig. 2). These uncertainty ranges consequently well include the expected signals of the two spikes at depths of about 48.07 and 50.85 m (Fig. 2). In addition, according to our model A chronology, the peak-to-peak separation of the two spikes is 51 years, and 47 years according to the model B chronology. These separations are again very close to the value of 48 years, representing the historically known time difference between the explosions of SN 1006 and the Crab Nebula.

Let us now examine the probability of the chance occurrence of two spikes in the uncertainty ranges of the expected SN signal dates with the expected relative age difference of ~48 years. We can easily estimate an upper limit of the probability by considering that *at least* one spike occurs in the uncertainty range (15 years) corresponding to the expected signal date (AD 1007) for SN 1006 and *at least* one spike occurs in the uncertainty range of the relative age difference, $51 - 47 = 4$ years. Since three spikes appeared during a ~200-year period, the frequency of occurrence of $NO_3^-$ spikes in the core is ~0.015 per year, so the probability of no spikes in a year is $q \approx$ 0.985. Thus, the upper limit of the probability of a chance occurrence can be calculated to be $(1 - q^{15}) \times (1 - q^4) \approx 1 \times 10^{-2}$, or 1 %. In this calculation, any number of spikes might occur anywhere; we did not exclude such a case of third spike occurring between the spikes at 48.07 and 50.85 m depth, for example. Also, for simplicity, we did not fix the location of the 15-year uncertainty range on the absolute dating axis (see Fig. 2). Hence, the actual probability should be much smaller than 1%. Thus, it is very unlikely that these two spikes occurred in these positions by chance. In other words, these two spikes are associated with SN 1006 and the Crab Nebula, respectively, with a confidence level much larger than 99%.

We observed a modulation with a period of about 10 years in the background trend of the $NO_3^-$ concentration profile (Fig. 2). This modulation could represent the 11-year solar cycle, as reported in $NO_3^-$ concentration profiles in other cores from Antarctica[11,25,26]. If this is the case, determination of the period of the modulation in our $NO_3^-$ data could be used to reinforce the validity of our core dating. We hence performed periodogram analyses on the $NO_3^-$ time-series data. Figure 4 shows the results in relation to the model B chronology, obtained by using the epoch folding method[27], which has a clear mathematical basis. Here, we have omitted the five data points larger than 0.4 μeq $L^{-1}$ that were spike components. The periodogram shows a peak at 10.7±0.3 (FWHM) years ($\chi^2 = 22.1$, corresponding to the 99.9 % confidence level; Fig. 4). With the model A chronology, the period becomes 11.6±0.3 years. This range, 10.7 to 11.6 years, is consistent with the known 11-year solar modulation. It is also in accordance with the solar periodicity of 9±1 (1σ) years[28] derived from radiocarbon measurements of tree rings within the 2σ uncertainty. These results strengthen our inferences about the MARs in MCA and hence further confirm that the model chronologies assumed here are appropriate.

The $NO_3^-$ concentration spikes interpreted to date to be associated with SPEs typically increase over 1 month or less[8]. Consequently, the SN 1006 candidate spike, which rises over ~2 years (Fig. 2), cannot be a SPE. On the other hand, the Crab Nebula candidate spike consists of only one data point; hence, its actual duration is not resolved in our data. However, as described above, SPEs are scarcely expected at around the middle of the 11[th] century, because solar activity was quiet at that time. Even the largest SPE-associated $NO_3^-$ spike[8] reported from Antarctica, if it occurred at the depth of the Crab Nebula candidate spike, would have a height (excess $NO_3^-$ concentration above the background level) only one third that of the Crab Nebula candidate spike. Therefore, we can reject the possibility that the Crab Nebula candidate spike was produced by a SPE. Moreover, using available knowledge of the radiation chemistry processes induced by γ-rays[29] and important neutral chemical reactions of nitrogen in the stratosphere[30], we estimated the order of magnitude of nitrogen oxide production under a simple assumption that the SN γ-ray energy is equipartitioned within an atmospheric column at the altitudes between 25 and 45 km. We found that SN 1006 could produce nitrogen oxides about one order of magnitude higher than that from solar radiation, and that the Crab Nebula of the same order of magnitude. From the energetics point of view, it is thus reasonable that the SN spikes with large amplitudes appear above the solar cycle background trend.

Finally, we conjecture that the third spike, which we dated to around AD 1060–1080 (Fig. 2), may be the signal of an unknown astronomical event. A giant burst from a soft γ-ray repeater could account for this spike. Alternatively, it could be the signal of a SN that was visible only from the Southern Hemisphere or one that exploded behind a dark cloud. Further extending our analysis to deeper and shallower ice cores would give fruitful information on galactic SN and solar activity histories, the dating of ice cores using SN signals as time markers, and even the effects of atmospheric circulation. Ionic measurements covering the past 2,000 years, including analyses of all known historical SNe and solar periodicity, is now in progress.

1. Rood, R. T., Sarazin, C. L., Zeller, E. J. & Parker, B. C. X- or γ-rays from supernovae in glacial ice. *Nature* **282**, 701–703 (1979).

2. Risbo, T., Clausen, H. B. & Rasmussen, K. L. Supernovae and nitrate in the Greenland Ice Sheet. *Nature* **294**, 637–639 (1981).

3. Herron, M. M. Impurity Sources of F$^-$, Cl$^-$, NO$_3^{3-}$, and SO$_4^{2-}$ in Greenland and Antarctic precipitation. *J. Geophys. Res.* **87**, 3052–3060 (1982).

4. Legrand, M. R. & Kirchner, S. Origins and variations of nitrate in South polar precipitation. *J. Geophys. Res.* **95**, 3493–3507 (1990).

5. Green, D. A. & Stephenson, F. R. On the footprints of supernovae in Antarctic ice cores. *Astroparticle Phys*. **20**, 613–615 (2004).

6. Dreschhoff, G. A. M. & Laird, C. M. Evidence for a stratigraphic record of supernovae in polar ice. *Advances in Space Res. (a COSPER publication)* **38**, 1307–1311 (2006).

7. Kameda, T., Motoyama, H., Fujita, S. & Takahashi, S. Temporal and spatial variability of surface mass balance at Dome Fuji, East Antarctica, by the stake method from 1995 to 2006. *J. Glaciol*. **54**, 107–116 (2008).

8. McCracken, K. G. *et al*. Solar cosmic ray events for the period 1561–1994 1. Identification in polar ice, 1561-1950. *J. Geophys. Res*. **106**, 21,585–21,598 (2001).

9. Siscoe, G. L. Evidence in the auroral record for secular solar variability. *Rev. Geophys. Space Phys.* **18**, 647–458 (1980).

10. Stothers, R. Giant solar flares in Antarctic ice. *Nature* **287**, 365–365 (1980).

11. Dreschhoff, G. A. M., Zeller, E. J. & Parker, B. C. Past solar activity variation reflected in nitrate concentrations in Antarctic ice. in *Weather and Climate Response to Solar Variations* (ed. McCormac, B. M.) 225–236 (Colorado Associated University Press, Boulder, 1983).

12. Cole-Dai, J. *et al*. A 4100-year record of explosive volcanism from an East Antarctica ice core. *J. Geophys. Res*. **105**, 24,431–24,441 (2000).

13. Palais, J. M., Germani, M. S. & Zielinski, G.A. Inter-hemispheric transport of volcanic ash from a 1259 A.D. volcanic eruption to the Greenland and Antarctic ice sheets. *Geophys. Res. Lett.* **19**, 801–804 (1992).

14. Siebert, L. & Simkin, T. Volcanoes of the world: an illustrated catalog of Holocene volcanoes and their eruptions. Smithsonian Institution, Global volcanism program digital information series, GVP-3, Large Holocene eruptions, http://www.volcano.si.edu/world/largeeruptions.cfm (2002–).

15. Langway, C. C. *et al*. A 10-century comparison of prominent bipolar volcanic events in ice cores. *J. Geophys. Res*. **100**, 16,241–16,247 (1995).

16. Hammer, C. U., Clausen, H. B., & Langway, C. C. 50,000 years of recorded global volcanism. *Climatic Change* **35**, 1–15 (1997).

17. Zielinski, G. A. *et al*. Record of volcanism since 7000B.C. from the GISP2 Greenland ice core and implications for the volcano-climate system. *Science* **264**, 948–952 (1994).

18. Watanabe, O. *et al*. A preliminary study of ice core chronology at Dome Fuji station, Antarctica. in *Proc. NIPR Symp. Polar Meteorol. Glaciol*. **11** (eds. Watanabe, O. *et al*.) 9–13 (National Institute of Polar Research, Tokyo, 1997).

19. Jones, P. B. & Mann, M. E. Climate over past millennia. *Rev. Geophys*. **42**, RG2002 (2004).

20. Narita, H. *et al*. Estimation of annual layer thickness from stratigraphical analysis of Dome Fuji deep ice core. in *Mem. Natl. Inst. Polar Res. Spec. Issue* **57** (eds. Shoji, H. & Watanabe, O.) 38–45 (National Institute of Polar Research, Tokyo, 2003).

21. Williams, M. B., Aydin, M., Tatim, C. & Saltzman, E. S. A 2000 year atmospheric history of methyl chloride from a South Pole ice core: Evidence for climate-controlled variability. *Geophys. Res. Lett*. **34**, L07811 (2007).

22. Khim, B.-K. & Bahk, J. J. Unstable climate oscillations during the late Holocene in the Eastern Bransfield Basin, Antarctic Peninsula. *Quaternary Research* **58**, 234–245 (2002).

23. Meese, D. A. *et al*.  The accumulation record from the GISP2 core as an indicator of climate change throughout the Holocene. *Science* **266**, 1680–1682 (1994).

24. Horiuchi, K. *et al*. Ice core record of [10]Be over the past millennium from Dome Fuji, Antarctica: A new proxy record of past solar activity and a powerful tool for stratigraphic dating. *Quaternary Geochronology* **3**, 253–261 (2008).

25. Zeller, E. J. & Parker, B. C. Nitrate ion in Antarctic firn as a marker for solar activity. *Geophys. Res. Lett*. **8**, 895–898 (1981).

26. Watanabe, K. *et al*.  Non-sea-salt sulfate and nitrate variations in the S25 core, near the coastal region, east Antarctica. *Polar Meteorol. Glaciol*. **13**, 64–74 (1999).

27. Davies, S. R. An Improved test for periodicity. *Mon. Not. R. astr. Soc*. **244**, 93–95 (1990).

28. Miyahara, H., Yokoyama, Y. & Masuda, K. Possible link between multi-decadal climate cycles and periodic reversals of solar magnetic field polarity. *Earth Planet. Sci. Lett.* **272**, 290–295 (2008).

29. Maetzing, H. Chemical kinematics of flue gas cleaning by irradiation with electrons. in *Advances in Chemical Physics* **80 (**eds. Prigogine, I. & Rice, S. A.**)**, 315–402 (Wiley-Interscience, New York, 1991).

30. Brasseur, G. P. & Solomon, S. *Aeronomy of the Middle Atmosphere, Third revised and enlarged Edition* (Springer, Dordrecht, 2005).

**Acknowledgements** We thank all members of the Dome Fuji drilling team and the JARE-42 party, and the participants of the science group working on the DF2001 core at NIPR in 2004. We are indebted to T. Kobayashi for the measurements of ionic concentrations used in this study. We are also grateful to K. Horiuchi, T. Kameda, S. Tsuneta, K. Masuda, S. Fujita, and H. Akiyoshi for comments and discussions. Y.M. and K.M. extend their special thanks to T. Ohata for introducing them to NIPR to launch this study. This work was supported in part by a Grant-in-Aid for Scientific Research from the Japan Society for the Promotion of Science, the General Cooperative Research in NIPR, and the Special Research Project in Basic Science in RIKEN.

Figure 1. Location of Dome Fuji station (77.2ºS, 39.4ºE; 3810 m ASL) at the highest point in east central Antarctica.

Figure 2. Profile of $NO_3^-$ concentrations in a portion of an ice core drilled in 2001 at Dome Fuji station, shown as a function of core depth (bottom axis) and our model chronologies (top axis). The blue open squares indicate replicated measurement results. The measurement error (±0.01 μeq L$^{-1}$) is shown only at 55.98 m depth (see Supplementary Information). The horizontal bars denote ranges within which SN 1006 and the Crab Nebula signals would be expected (with widths of 15 and 19 years, respectively), as determined from our dating uncertainties. The vertical, blue dashed lines indicate the positions of AD 1007 (SN 1006) and AD 1055 (Crab Nebula) in relation to model A chronology, and the green dashed lines their positions according to model B chronology.

Figure 3. Assumed MARs yielding our two model chronologies. The blue solid line indicates the MARs (model A) obtained from the identified volcanic eruptions. The green dashed line shows the MARs (model B) derived from the $\delta^{18}O$ measurements in the DF2001 core. Values in parentheses are the corresponding dates AD. The maximum MAR value in model B was obtained by conserving the total amount of ice that accumulated between the AD 639 and 1168 volcanic time markers.

Figure 4. A periodgram of the row data series of $NO_3^-$ concentrations (excluding the five data points exceeding 0.4 μeq L$^{-1}$), obtained by the epoch folding method[27], with our model B chronology. Each period cycle was divided into 7 phase bins.

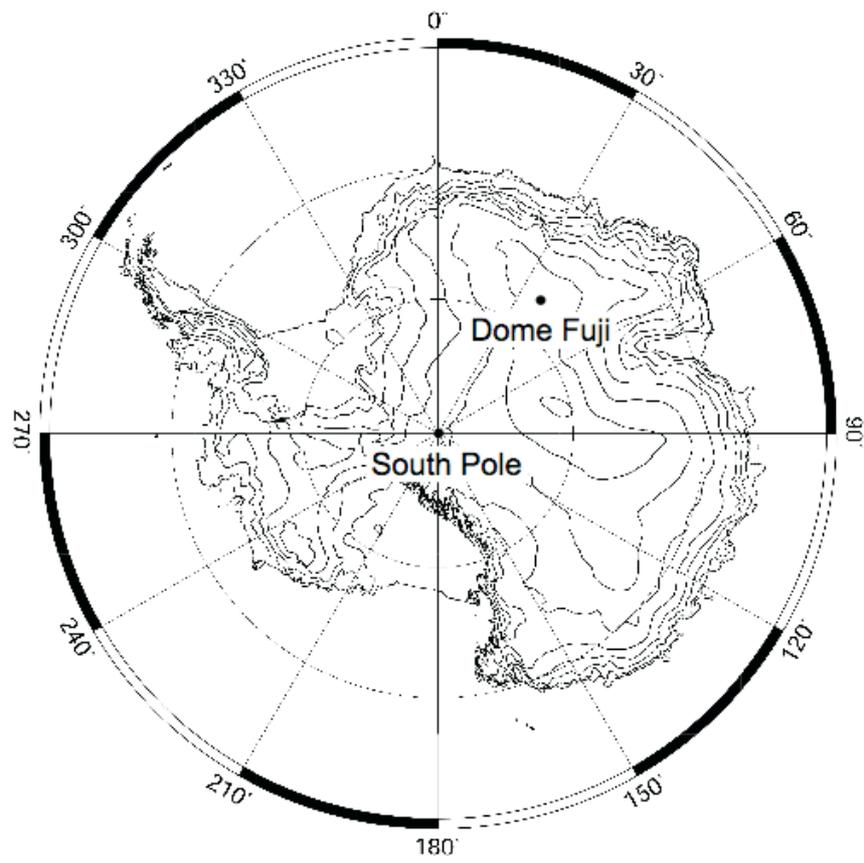

Fig. 1. Location of Dome Fuji station.

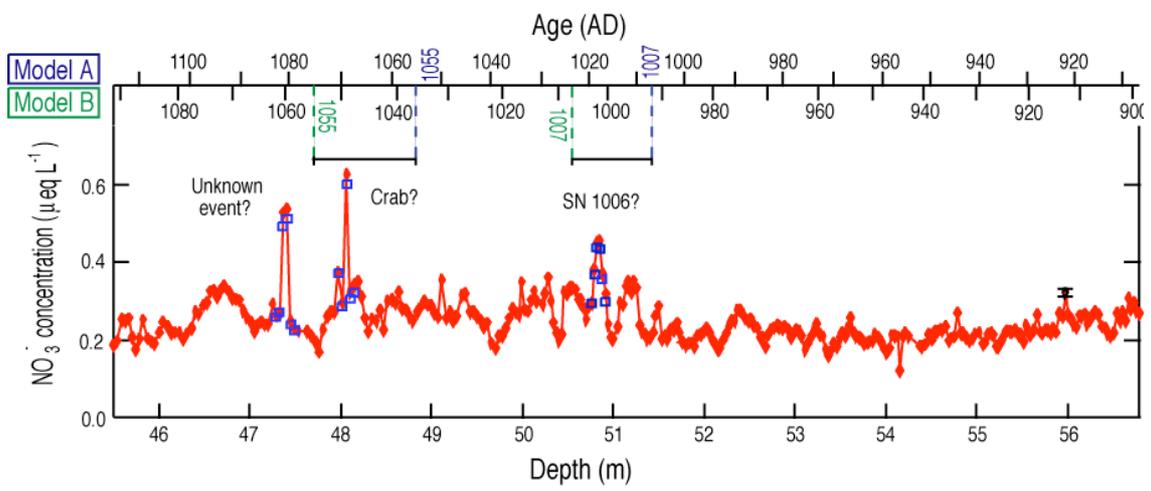

Fig. 2. Profile of $NO_3^-$ concentrations.

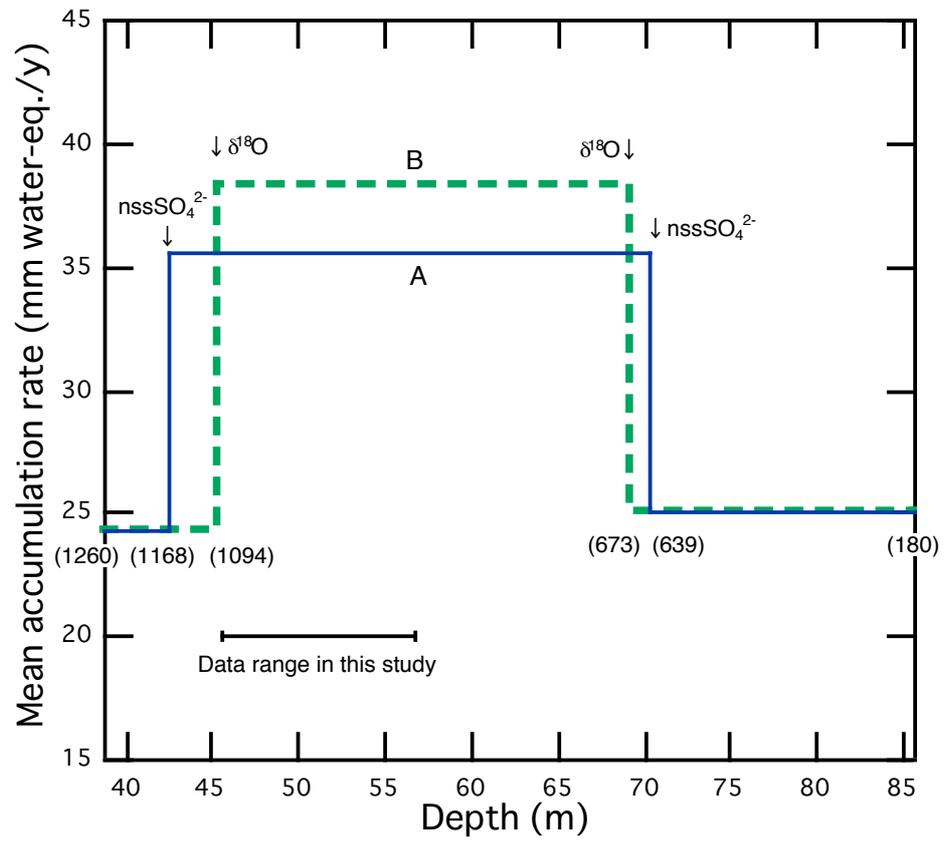

Fig. 3. Assumed MARs.

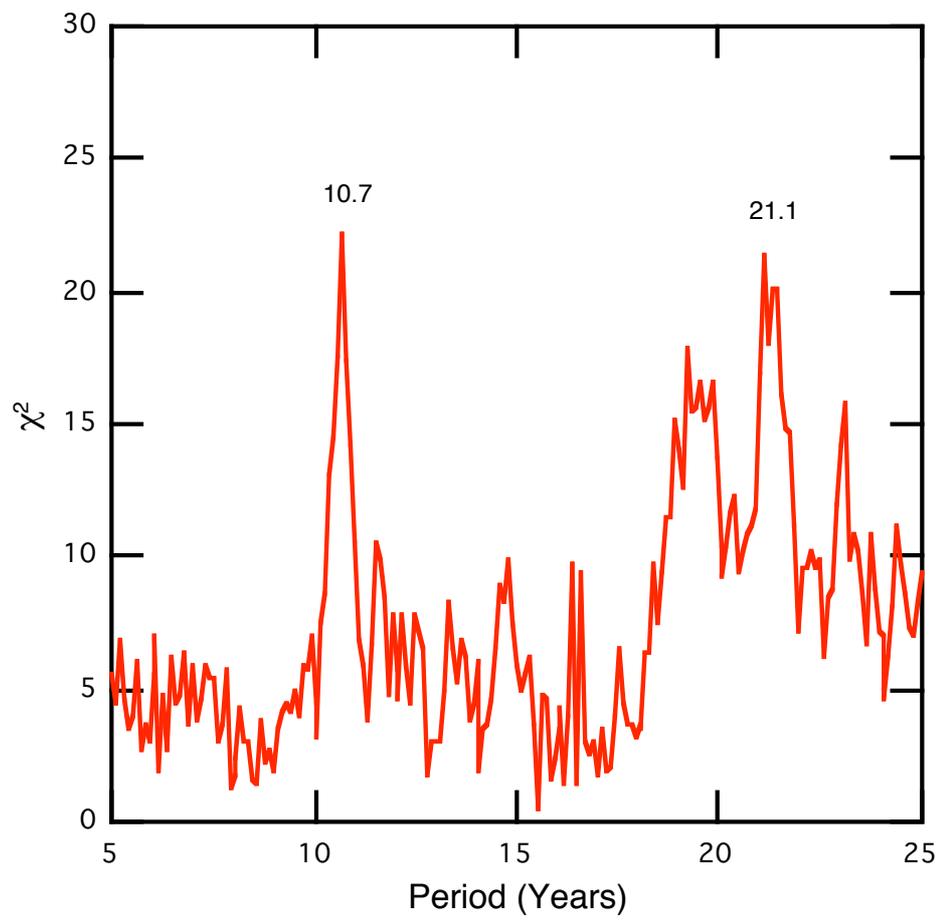

Fig. 4. A periodogram of $NO_3^-$ concentrations.

**Supplementary Methods (Supplementary Information)**

We analysed the concentrations of both anions and cations with the Dionex 500 high-performance ion chromatography (IC) system at the NIPR (Japan). Figure S1 shows the relationship between the signal intensity (area) observed with IC and the concentration of $NO_3^-$ ion in the calibration solutions. Four different solutions were measured twice; the reproducibility was quite good (in Fig. S1, it is difficult to distinguish the two blue data points plotted for the same concentration). Furthermore, we obtained an extremely straight calibration line, indicating a measurement error within ±0.01 μeq L$^{-1}$, or 3–4% against the background level. Since this error is very small, it is concluded that the sample variance of our $NO_3^-$ ionic data was governed not by the measurement error but by other causes.

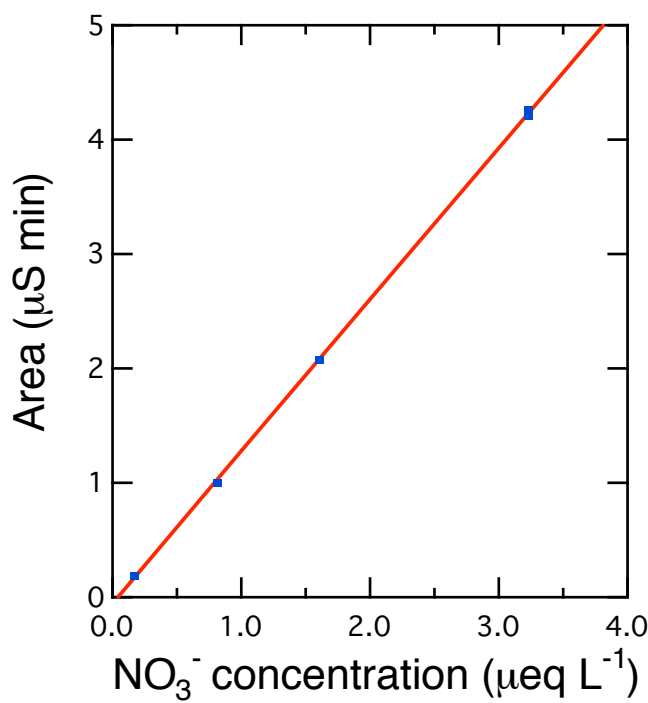

Figure S1. A typical calibration line for the $NO_3^-$ measurements with IC, relevant to our study.